\documentclass[aps,prd,amsmath,amssymb,citeautoscript,superscriptaddress]{revtex4-2}
\usepackage{graphicx}
\graphicspath{ {./images/} }
\usepackage{dcolumn}
\usepackage{bm}
\usepackage{accents}
\usepackage{amsmath}
\usepackage{comment}
\usepackage{mathtools}

\newcommand{\aap}{Astron.\ Astrophys.}

\newcommand{\apjl}{Astrophys.\ J.\ Lett.}

\begin{document}

\title{Identifying the Event Horizons of Parametrically Deformed Black-Hole Metrics}

\author{Dirk Heumann}
\affiliation{Department of Astronomy and Steward Observatory, University of Arizona, 933 N. Cherry Ave., Tucson, AZ 85721, USA}

\author{Dimitrios Psaltis}
\affiliation{Steward Observatory and Department of Astronomy, University of Arizona, 933 N. Cherry Ave., Tucson, AZ 85721, USA}


\begin{abstract}
Recent advancements in observational techniques have led to new tests of the general relativistic predictions for black-hole spacetimes in the strong-field regime. One of the key ingredients for several tests is a metric that allows for deviations from the Kerr solution but remains free of pathologies outside its event horizon. Existing metrics that have been used in the literature often do not satisfy the null convergence condition  that is necessary to apply the strong rigidity theorem and would have allowed us to calculate the location of the event horizon by identifying it with an appropriate Killing horizon. This has led earlier calculations of event horizons of parametrically deformed metrics to either follow numerical techniques or simply search heuristically for coordinate singularities. We show that several of these metrics, almost by construction, are circular. We can, therefore, use the weak rigidity and Carter's rotosurface theorem and calculate algebraically the locations of their event horizons, without relying on expansions or numerical techniques. We apply this approach to a number of parametrically deformed metrics, calculate the locations of their event horizons, and place constraints on the deviation parameters such that the metrics remain regular outside their horizons. We show that calculating the angular velocity of the horizon and the effective gravity there offers new insights into the observational signatures of deformed metrics, such as the sizes and shapes of the predicted black-hole shadows.
\end{abstract}


\maketitle

\section{Introduction}

General Relativistic predictions for the spacetimes of black holes are now being tested against observations of gravitational waves during the coalescence of stellar-mass black holes~\cite{Abbott2016,*Abbott2019,*Abbott2021}, the detection of post-Newtonian effects in the orbits of stars around Sgr~A* in the center of the Milky Way~\cite{Hees2017,*Amorim2019,*Do2019,*Abuter2020}, as well as imaging observations of supermassive black holes with horizon-scale resolution~\cite{PaperI,*SgrAPaperI,*SgrAPaperVI,Psaltis2020}. Performing these tests requires a framework that allows for deviations from these General Relativistic predictions. One of the ingredients of such a framework is a model for the equilibrium spacetime of the compact object that is allowed to be parametrically different from the Kerr spacetime.

The no hair theorem in General Relativity ensures that the only black-hole spacetime that is stationary, axisymmetric, asymptotically flat, free of pathologies, and a solution to the Einstein field equations is the one described by the Kerr metric~\cite{Israel1967,*Israel1968,*Carter1968,*Carter1971,*Hawking1972,*Price1972a,*Price1972b,*Robinson1975}. (For astrophysical black holes, we do not consider the additional degrees of freedom introduced by an electric charge.) Introducing any deviations from the Kerr metric, therefore, requires that one of these basic assumptions for the spacetime is allowed to be violated. Early attempts considered metrics that are Ricci flat, i.e., metrics that are solutions to the Einstein field equations but describe either naked singularities~\cite{Gair2008} or have other pathologies such as closed time-like curves~\cite{Glampedakis2006,Johannsen2013a}.
Because the presence of pathologies outside the horizons often precludes the calculation of observable predictions, most recent attempts have instead abandoned the assumption of Ricci flatness and either introduce parametric deviations in a manner that is agnostic to the underlying theory~\cite{Collins2004,Johannsen2011,Johannsen2013b,Vigeland2011,Cardoso2014,Rezzolla2014,*Konoplya2016} or that are specific solutions to modified field equations~\cite{Kanti1996,*Kanti1998,*Yunes2009,*Yunes2011,*Yagi2012,*Barausse2013,*Berti2015,*Ayzenberg2014,*McNees2016,*Silva2018,*Antoniou2018a,*Antoniou2018b,*Doneva2018}. Invariably, all these vacuum spacetimes contain pathologies but, as long as they reside inside event horizons, they do not hamper the calculation of observables. Identifying the presence of event horizons and calculating their locations is, therefore, critical for assessing whether a parametric non-Kerr spacetime is suitable for tests of gravity with black-hole observations.

The event horizon of an asymptotically flat spacetime is a continuous null surface that separates spacetime locations from which a future null geodesic can reach future null infinity from those that cannot. A variety of techniques have been developed for event-horizon finding, primarily to address the needs of the numerical relativity community~\cite{Thronburg2007}. Albeit not complicated, especially in the case of the stationary and axisymmetric spacetimes usually considered, these techniques involve the solution of differential equations, which often can be achieved only numerically. Such techniques have been employed in Ref.~\cite{Johannsen2013a} to find the horizons of several non-Kerr spacetimes. 

In General Relativity, one can use Hawking's strong rigidity theorem to connect the global concept of an event horizon to the local concept of a Killing horizon, which is easier to calculate~\cite{Hawking1972}. The strong rigidity theorem requires that the Ricci tensor of the spacetime satisfies the null convergence condition $R_{\mu\nu}l^\mu l^\nu\ge 0$, for all null vectors $l^\mu$~\cite{Chrusciel1996,Chrusciel2012}. In General Relativity, this condition is closely connected to the weak energy condition~\cite{Curiel2014} since
\begin{equation}
    R_{\mu\nu}l^\mu l^\nu = 8 \pi (T_{\mu\nu} - \frac{1}{2} T g_{\mu\nu}) l^{\mu} l^{\nu} = 8 \pi T_{\mu\nu} l^{\mu} l^{\nu}\ge 0\;,
\end{equation}
which is expected to be satisfied. For spacetimes that do not obey the Einstein field equations, however, the null convergence condition is not necessarily satisfied and, therefore, one cannot apply the strong rigidity theorem to calculate the event horizon.

The more powerful, albeit less general, weak rigidity theorem~\cite{Carter1968}, which shows that the rotosurface is a Killing Horizon, and Carter's rotosurface theorem \cite{Carter2010}, which says that the rotosurface is the event horizon, connect the event horizon of a spacetime with a Killing horizon without making any assumptions regarding the underlying field equations, if the spacetime is stationary, axisymetric, and circular (see Ref.~\cite{Heusler1996} for an introduction to rigidity theorems). A spacetime is circular if it has two Killing vectors $k$ and $m$ and
\begin{eqnarray}
k^{[\mu} m^{\nu} \nabla^{\lambda} m^{\rho]} = 0 \\
m^{[\mu} k^{\nu} \nabla^{\lambda} k^{\rho]} = 0\;.
\end{eqnarray}
We use here the term ``circular'' for any gravity theory and for vacuum spacetimes that satisfy these conditions, even though the term was originally introduced for General Relativistic spacetimes in the presence of matter, which obey the circularity condition only if the matter field is in circular orbits, i.e., with no meridian motions~\cite{Gourgoulhon2010}. 

When expressed in adapted coordinates $(t,r,\theta,\phi)$, a circular spacetime can be separated into two sub-manifolds: the $M_{(t\phi)}$ one spanned by the orbits of the two Killing vectors and the orthogonal manifold $M_{(r\theta)}$, whose tangent vectors are orthogonal to the tangent vectors of the orbital sub-manifold $M_{(t\phi)}$. The orthogonality of the tangent vectors of $M_{(r\theta)}$ and $M_{(t\phi)}$ allows for the covariant metric (defined on the tangent space) to be decomposed into a metric on the tangent vectors of $M_{(t\phi)}$ and a metric on the tangent vectors of $M_{(r\theta)}$, with no cross terms. Furthermore, the metric components in a separable structure only depend on the non-ignorable coordinates $r,\theta$. 

Almost by construction, the vast majority of parametrically deformed Kerr metrics obey the circularity condition, as do many of the black-hole metrics that are solutions to modified field equations. This property allows us to find their event horizons, without resorting to complex integrations of differential equations. In Section~II, we demonstrate this approach on two widely used parametric metrics, which we show are indeed circular and use this property to calculate the locations of their event horizons. In Section~III, we evaluate the circularity condition for two spacetimes that are solutions to modified field equations, show that one satisfies it and the other does not, and discuss the applicability of the above theorem. We offer our brief conclusions in Section~IV. 

\section{Parametrically Deformed Black-Hole Metrics}

In this section, we evaluate the validity of the circularity condition for two metrics that are parametrically different from Kerr and use Carter's rotosurface theorem and the weak rigidity theorem to calculate the locations of their event horizons.

\subsection{The Johannsen-Psaltis metric}

The first metric we will explore is the one introduced by Johannsen \& Psaltis~\cite{Johannsen2011,Johannsen2013b} (hereafter the JP metric) and further developed in, e.g., Refs.~\cite{Cardoso2014}. This separable metric was developed to be free of pathologies, while allowing for parametric deviations from the Kerr metric and for the existence of three integrals of motion for the trajectories of particles and photons. It is generated from the static Schwarzschild metric by adding higher order terms in $1/r$, which maintains staticity, and then performing a Newman-Janis-transformation, which loses staticity, when real $r$-dependent terms are replaced by suitable combinations of complex $r$ and $\overline{r}$ terms. This procedure, however, does maintain circularity, as only radial and not azimuthal terms are modified by the replacement prescription of $r$.

The general form of the JP metric in Boyer-Lindquist like coordinates is
\begin{eqnarray}
ds^2  =&-&\frac{\tilde{\Sigma}(\Delta-a^2 A^2_2 \sin^2 \theta)}{[(r^2+a^2)A_1-a^2 A_2 \sin^2 \theta]^2} dt^2 - \frac{2 a [(r^2+a^2) A_1 A_2 -\Delta] \tilde{\Sigma} \sin^2 \theta}{[(r^2+a^2) A_1 -a^2 A_2 \sin^2 \theta]^2} dt d\phi \nonumber \\
&+&  \frac{1}{A_5} \frac{\Sigma}{\Delta} dr^2+  \frac{\tilde{\Sigma} \sin^2 \theta [(r^2+a^2)^2 A_1^2 -a^2 \Delta \sin^2 \theta]}{[(r^2+a^2)A_1-a^2 A_2 \sin^2 \theta]^2} d\phi^2 + \Sigma d\theta^2\;,
\end{eqnarray}
with 
\begin{eqnarray}
A_1&=&1+\frac{\alpha_{13}}{r^3}+{\cal O}(r^{-4})\;,\nonumber\\
A_2&=&1+\frac{\alpha_{22}}{r^2}+{\cal O}(r^{-3})\;, \nonumber\\
A_5&=&1+\frac{\alpha_{52}}{r^2}+{\cal O}(r^{-3})\;,\nonumber\\
\tilde{\Sigma}&=&r^2+a^2 \cos^2 \theta + \frac{\epsilon_3}{r}+{\cal O}(r^{-2})
\end{eqnarray}
and the usual Kerr definitions for $\Delta = r^2 - 2  r +a^2 $ and  $\Sigma = r^2+a^2 \cos^2 \theta $. The JP metric reduces to the standard Kerr metric for $\alpha_{ij} = 0$ and $\epsilon_3 = 0$. Unless otherwise specified, we set $G=c=M=1$, where $G$ is the gravitational constant, $c$ is the speed of light, and $M$ is the mass of the compact object, as measured with Keplerian orbits at large distances.

The JP metric, in general, does not obey the null convergence condition. To verify this, we take the limit $M\ll 1$ and set the spin and all deviation coefficients $\alpha_{ij}$ of the JP metric to zero. The metric becomes diagonal with $g_{rr} = -g_{tt} = 1 + \frac{\epsilon_3}{r^3}$. This means that the 4-vector $l= \partial_t + \partial_r$ is null. We obtain
\begin{equation}
R_{\mu \nu} l^\mu l^\nu = R_{tt} + R_{rr} = \frac{3 \epsilon_3}{r^2 (r^3 + \epsilon_3)} \left[1-\frac{10 r^3 + \epsilon_3}{2(r^3 + \epsilon_3)}\right]
\xrightarrow[r\gg \vert\epsilon_3\vert^{1/3}]{}-\frac{12\epsilon_3}{r^5}\;.
\end{equation}
This last expression can be positive or negative at large distances from the horizon, depending on the sign of the parameter $\epsilon_3$. As a result, the null convergence condition is not satisfied and Hawking's strong-rigidity theorem cannot be applied to identify the event horizon of the JP metric with a Killing horizon. \\

We can demonstrate explicitly, however, that the JP metric is circular and, therefore, can apply the weak rigidity theorem and the rotosurface theorem. In adapted Boyer-Lindquist coordinates, as written above, the JP metric has two Killing vectors given by $k= \partial_t$ and $m= \partial_{\phi}$, because the metric coefficients are independent of the coordinates $t$ and $\phi$, i.e., $\partial_{t} g_{\mu\nu} = \partial_{\phi} g_{\mu\nu} = 0$. 
In order to assess the circularity of the JP metric, we write 
\begin{eqnarray}
k^{[\mu} m^{\nu} \nabla^{\lambda} m^{\rho]} = k^{t}  m^{\phi} ( \nabla^{r} m^{\theta} - \nabla^{\theta} m^{r}) = g^{r r} \nabla_{r} m^{\theta} - g^{\theta \theta} \nabla_{\theta} m^{r} = g^{r r} \Gamma^{\theta}_{r \phi}-g^{\theta \theta} \Gamma^{r}_{\theta \phi} = 0\;,
\end{eqnarray}
where we used the fact that the only non-zero components of the two Killing vectors are $k^t = 1$ and $m^{\phi}=1$ and that $\Gamma^{\theta}_{r \phi}= \Gamma^{r}_{\theta \phi} =0$. A similar argument holds for $m^{[\mu} k^{\nu} \nabla^{\lambda} k^{\rho]} = 0$. The JP metric is, therefore, circular. Note that, even though we have shown explicitly the circularity property of the JP metric, the same conclusion can be reached for any spacetime that follows the general form  of the Lewis-Papapetrou metric. 

According to the weak rigidity and the rotosurface theorems, the event horizon of the JP metric is also a Killing horizon. In order to calculate the location of the latter, we first evaluate the quantity 
\begin{equation}
{\cal N} = -(k \wedge m \vert k \wedge m)\;,
\end{equation}
which in adapted coordinates is the negative determinant of the $t,\phi$-part of the metric, i.e.,
\begin{equation}
{\cal N} = -(k \wedge m \vert k \wedge m) = -[(k \vert k) (m \vert m) - (k \vert m)^2] \overset{*}{=} -(g_{tt} g_{\phi \phi} - g^2_{t\phi}) = -\det g_{(t\phi)}\;.
\end{equation}
Here, the symbol $\overset{*}{=}$  denotes the values of the scalar quantities $(k \vert k)$,$(k \vert m)$and $(m \vert m)$ in Boyer-Lindquist coordinates.
On the Killing horizon,  ${\cal N}=0$ ~\cite{Heusler1996}. Therefore, the roots of ${\cal N}$ trace the Killing horizon of this circular spacetime, which is identical to its event horizon~\footnote{Ref.~\cite{Johannsen2013b} explicitly calculated the location of the event and the Killing horizons and showed that they coincide, without assigning this coincidence to the validity of the weak rigidity theorem.}.

In order to calculate the location of the horizon in the JP metric, it is easier first to express the latter in terms of rational polynomial coordinates, i.e.,
\begin{eqnarray}
ds^2  =&-&\frac{\tilde{\Sigma}[\Delta-a^2 A^2_2 (1-z^2)]}{[(r^2+a^2)A_1-a^2 A_2 (1-z^2)]^2} dt^2 - \frac{2 a [(r^2+a^2) A_1 A_2 -\Delta] \tilde{\Sigma} (1-z^2)}{[(r^2+a^2) A_1 -a^2 A_2 (1-z^2)]^2} dt d\phi \nonumber \\
&+&\frac{1}{A_5} \frac{\Sigma}{\Delta} dr^2+  \frac{\tilde{\Sigma} (1-z^2) [(r^2+a^2)^2 A_1^2 -a^2 \Delta (1-z^2)]}{[(r^2+a^2)A_1-a^2 A_2 (1-z^2)]^2} d\phi^2 +  \frac{\Sigma}{(1-z^2)} dz^2\;,
\end{eqnarray}
with the various coefficients expressed as before. The determinant ${\cal N}$ in the JP metric is
\begin{eqnarray}
{\cal N} &=& \frac{r^4 \ \Delta \ (1-z^2) (r^3+a^2 r z^2+\epsilon_3)^2 }{[r^5 +a^2 r^3 z^2 + \alpha_{13} \ (r^2 + a^2) - \alpha_{22} \ a^2 r (1-z^2)]^2}  \nonumber \\
&=&  \frac{r^6 \ \Delta \ \tilde{\Sigma}^2 \ (1-z^2)}{[r^3 \ \Sigma + \alpha_{13} \ (r^2 + a^2) - \alpha_{22} \ a^2 r (1-z^2)]^2} \equiv \Delta \ h(r,z)
\end{eqnarray}
and is zero for $r=0$, $\Delta = 0$, and $\tilde{\Sigma} = 0$. The outer event horizon is located at $\Delta = 0$, i.e., $r=r_+ = 1 + \sqrt{1 - a^2}$, just as in the case of the Kerr metric. The location of the event horizon does not depend on the choice of the deviation parameters.

A second condition for the applicability of Carter's rotosurface theorem is that the metric remains regular outside the event horizon, i.e., that $\det g_{\mu\nu} < 0$ and never blows up. This condition was explored in detail in Ref.~\cite{Johannsen2013b}, who used it to place  bounds on the possible values of the deviation parameters.

The weak rigidity theorem allows us to calculate a few additional properties of the JP metric (see Ref.~\cite{Heusler1996} for details). The angular velocity of the outer event horizon is equal to
\begin{eqnarray}
\Omega_{\rm H} &=& -\frac{(k|m)}{(m|m)}\vert_{EH} =-\frac{(k|k)}{(k|m)}\vert_{EH} \overset{*}{=}-\left.\frac{g_{tt}}{g_{t\phi}} \right\vert_{r_+} = -\left.\frac{g_{t\phi}}{g_{\phi\phi}} \right\vert_{r_+} \nonumber \\ 
&=& \frac{a}{r_+^2+a^2}\left(\frac{A_2}{A_1}\right)= \frac{a}{r_+^2+a^2} \left(1+\frac{\alpha_{22}}{r_+^2}\right)
\left(1+\frac{\alpha_{13}}{r_+^3}\right)^{-1}\;.
\end{eqnarray}
This is constant along the horizon, as expected from the theorem, which states that the horizon rotates rigidly for circular metrics. We note that $\Omega_{\rm H}$ does not depend on the other two parameters $\epsilon_3$ and  $\alpha_{52}$. The first of these parameters enters the term $\tilde{\Sigma}$, which is a common factor for $g_{tt}, g_{t\phi}$, and $g_{\phi\phi}$ and, therefore, drops out of the calculation of $\det g_{(t\phi)} = 0$. The second of these parameters enters only the $g_{rr}$ component and, therefore, does not impact the calculation of $\det g_{(t\phi)}$.

The Killing vector $l^\nu$ of the JP metric that corresponds to the Killing horizon is equal to
\begin{equation}
l^\nu = k^\nu + \Omega_{\rm H}\  m^\nu = k^\nu + \frac{a}{r_+^2+a^2} \left(1+\frac{\alpha_{22}}{r_+^2}\right)
\left(1+\frac{\alpha_{13}}{r_+^3}\right)^{-1} \ m^\nu\;.
\end{equation}
Finally, the surface gravity on the outer event horizon for the JP metric is 
\begin{equation}
\kappa = \left[- \frac{1}{4} (dl|dl)  \right]^{1/2} = \left[- \frac{1}{2} \nabla_{\mu} l_{\nu} \nabla^{\mu} l^{\nu} \right]^{1/2}  = \frac{r_+-1}{r_+^2 + a^2} \left[1+\left(\frac{\alpha_{13}}{r_+^3}\right)^2\right]^{-1/2}\;,
\end{equation}
which is also clearly constant along the horizon and vanishes for an extremal black hole ($r_+=1$). The surface gravity only depends on the deviation parameter $\alpha_{13}$ and not on $\alpha_{22}$. It defaults to the value $(r_+-1)/(r_+^2 + a^2)$ for the Kerr metric ($\alpha_{13}=0$) and the value $1/4$ for the Schwarzschild metric ($a=0$, $\alpha_{13}=0$). \\

It is interesting to note that, for the Kerr metric, the sum $\Omega_H^2 + \kappa^2$ evaluated on the event horizon is a function only of the horizon radius $r_+$, i.e.,
\begin{equation}
\Omega_H^2 + \kappa^2 = (\frac{a}{2r_+})^2+(\frac{r_+ -1}{2r_+})^2 = \frac{\Delta + 1}{4 r_+^2} = \frac{1}{4 r_+^2}
\end{equation}
This is not true any longer for the JP metric, for which the $\Omega_H^2 + \kappa^2$ evaluated on the event horizon depends on both $r_+$ and the parameter $\alpha_{22}$. 

\subsection{The Rezzolla-Zhidenko metric}

The RZ metric approach introduces the functions $N$, $W$, $K$, $\tilde{\Sigma}$, and $B$ such that the resulting metric takes the form
\begin{equation}
ds^2 = -\frac{(N^2-W^2 \sin^2 \theta)}{K^2} dt^2 - 2 W r \sin^2 \theta dt d\phi + K^2 r^2 \sin^2 \theta d\phi^2 + \tilde{\Sigma} (\frac{B^2}{N^2} dr^2 + r^2 d\theta^2)
\end{equation}
with suitable restrictions on the a priori arbitrary functions to make the metric asymptotically flat and to identify the mass parameter and angular momentum of a rotating black hole. The functions $N,W,K,B, \tilde{\Sigma}$ are specified to depend only on the coordinates $r$ and $\theta$. This ensures that $k=\partial_t$ and $m=\partial_\phi$ remain Killing vectors. 

To lowest order in the deviation parameters, the various parametric functions become (see Appendix~A of Ref.~\cite{Konoplya2016})
\begin{eqnarray}
\label{eq:a2}
N^2(r,\theta) &=& \left(1-\frac{r_0}{r}\right)\left[1-\frac{2-r_0}{r}
+ \frac{a^2+r_0^2-2r_0}{r^2} +
\frac{a_{01}\,r_0^3}{r^3}\right] + \left[\frac{a_{20} r_0^3}{r^3} +
\frac{a_{21}\,r_0^4}{r^4} + \frac{k_{21}}{1+\frac{k_{22}(1-\frac{r_0}{r})}{1+k_{23}(1-\frac{r_0}{r})}} \frac{r_0^3}{r^3} \right]\cos^2\theta\,,\nonumber\\
\label{eq:a3}
B(r,\theta) &=& 1 + \frac{b_{01}\,r_0^2}{r^2} +
\frac{b_{21}\,r_0^2}{r^2}\cos^2\theta\,,\\
\tilde{\Sigma}(r,\theta) &=& 1 +
\frac{a^2}{r^2}\cos^2\theta\,,\\
\label{eq:a4}
W(r,\theta) &=& \frac{1}{\Sigma(r,\theta)}\left[\frac{2 a}{r^2} +
  \frac{w_{01}\,r_0^3}{r^3} + 
  \frac{w_{21}\,r_0^3}{r^3}\cos^2\theta\right]\,,\\
\label{eq:a5}
  K^2(r,\theta) &=& 1 + \frac{a W(r,\theta)}{r} +
\frac{1}{\Sigma(r,\theta)}\left[\frac{a^2}{r^2} +
  \frac{k_{21}}{1+\frac{k_{22}(1-\frac{r_0}{r})}{1+k_{23}(1-\frac{r_0}{r})}} \frac{r_0^3}{r^3}\cos^2\theta\right]\,,
\end{eqnarray}
where we have set some coefficients to the values dictated by various requirements regarding the asymptotic behavior of the metric and regularity conditions near the horizon. Note that the continued fraction in $N^2$ and $K^2$ with $k_{22} = - a^2/r_0^2$ and $k_{23} = a^2/r_0^2$ is necessary to reproduce the Kerr metric.\\

The determinant of the $t$,$\phi$-part of the metric is 
\begin{equation}
{\cal N} =  N^2 r^2 \sin^2 \theta
\end{equation}
and, therefore, the locations of the Killing and event horizons can be found by the requirement $N^2=0$. We will first explore the location of the horizon in spherical symmetry, i.e., by setting $a=0$ and choosing appropriate values for the other parameters such that there is no dependence of $N^2$ on the polar angle $\theta$. This is the metric originally introduced in Ref.~\cite{Rezzolla2014}, with $\alpha_1\equiv \alpha_{01}$.

\begin{figure}
\centering
\includegraphics[width=.47\textwidth]{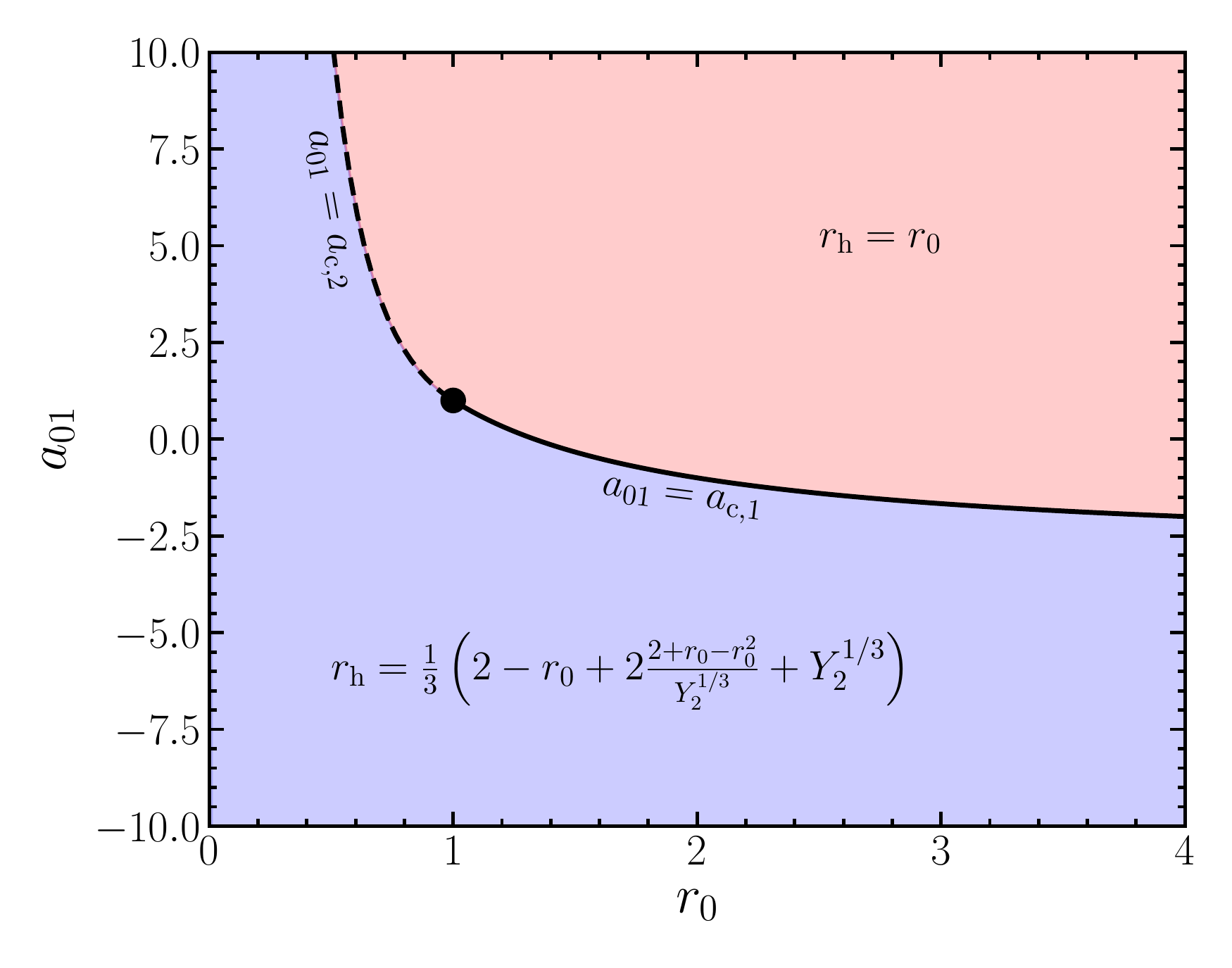}
\includegraphics[width=.47\textwidth]{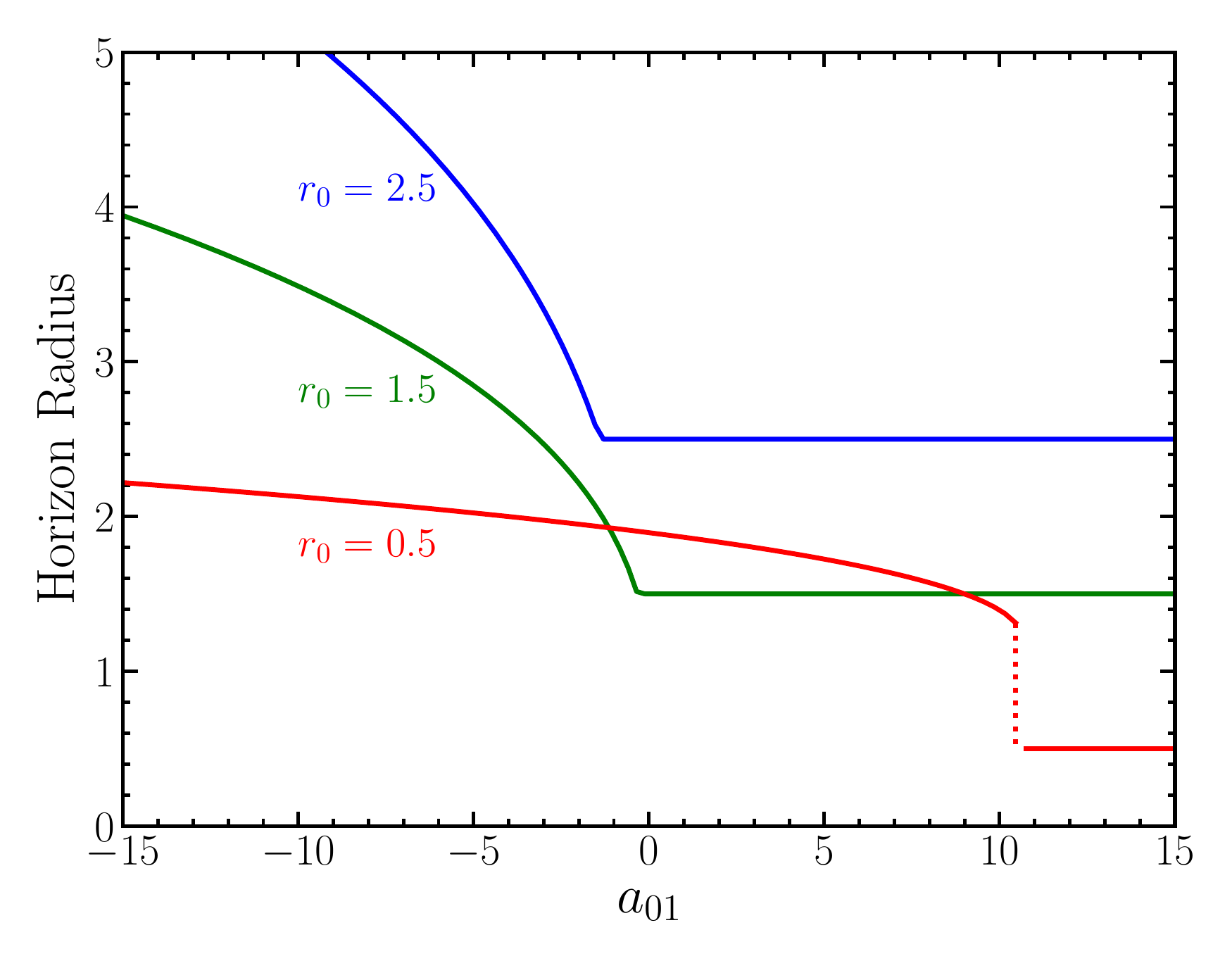}
\caption{{\em (Left)\/} Distinct regions in a cross section of the parameter space of the  spherically symmetric RZ metric for the calculation of the radius of the event horizon. The two expressions for the critical curve to the left and to the right of the solid circle are defined in the main text. The dashed segment of the curve signifies the fact that the horizon radius changes discontinuously there. {\rm (Right)\/} The dependence of the horizon radius on the deviation parameter $a_{01}$ of the spherically symmetric RZ metric, for different values of the parameter $r_0$. The change in the slope occurs on the critical curve shown on the left panel. When $r_0<1$, the change is discontinuous.}
\label{fig:rz_spher}
\end{figure}

By solving the algebraic equation ${\cal N}=0$ and exploring the relative ordering of the solutions, we find, contrary to the result reported in Ref.~\cite{Konoplya2016}, that the outer horizon occurs either at
\begin{equation}
    r_{\rm h,1}=r_0
\end{equation}
or at
\begin{equation}
    r_{\rm h,2}=\frac{1}{3}\left(2-r_0
    +2\frac{2+r_0-r_0^2}{Y_2^{1/3}}+Y_2^{1/3}\right)\;,
    \label{eq:rh_RZ_sph2}
\end{equation}
where
\begin{eqnarray}
Y_2&=&\frac{1}{2}\left\{Y_1+
\left[32(r_0-2)^3(r_0+1)^3 +Y_1^2\right]^{\frac{1}{2}}\right\}\nonumber\\
Y_1&=&16+12r_0-24 r_0^2+7 r_0^3 - 27 a_{01} r_0^3\;,
\end{eqnarray}
depending on the magnitudes of the parameters $r_0$ and $a_{01}$. In particular, when $r_0\ge 1$, the horizon is at $r_{\rm h,1}$ for 
\begin{equation}
    a_{01}\ge a_{\rm c,1}=(4-3 r_0)/r_0
\end{equation} 
and connects continuously to $r_{\rm h,2}$ for $a_{01}<a_{\rm c,1}$. On the other hand, when $r_0<1$,  the horizon remains at $r_{h,1}$ for large values of $a_{01}$ but jumps discontinuously to $r_{h,2}$ when
\begin{equation}
    a_{01}\le a_{\rm c,2}\equiv\frac{(2-r_0)}{27 r_0^3} \left[8+4 \sqrt{4+2r_0-2r_0^2}+r_0 \left(10-7r_0+4 \sqrt{4+2r_0-2r_0^2}\right)\right]\;.
\end{equation}
(Note that the expression for $r_{\rm h,2}$ is real when this last condition is satisfied but needs to be evaluated with care, as it involves the cancellation  of two imaginary terms of equal magnitude but opposite sign). Of course, these conditions will be modified if terms of higher order, i.e., $a_{02}$ and higher, are included. Figure~\ref{fig:rz_spher} shows the distinct regions of the $(r_0,a_{01})$ parameter space for the  calculation of the horizon radius, as well as two examples of the dependence of the horizon radius on the  parameter  $a_{01}$. 

The second condition for the applicability of Carter's rotosurface theorem is that the determinant of the metric is negative everywhere, i.e.,
\begin{equation}
\det g_{\mu\nu} = - B^2 \tilde{\Sigma}^2 r^4 \sin^2 \theta <0\;.
\end{equation}
This condition ensures that the metric is everywhere Lorentzian. The determinant of the metric is manifestly negative unless the coefficients of the $B$ function are such that $B=0$ somewhere outside the outer event horizon. For a spherically symmetric spacetime and keeping only the lowest-order deviation parameters, this implies that, in order for $B$ to not be zero when $r>r_{\rm h}$ (see eq.~[\ref{eq:a3}]), the $b_{01}$ parameter needs to satisfy
\begin{equation}
    b_{01}>-\left(\frac{r_h}{r_0}\right)^2\;.
\end{equation}
Because, as we have seen above, the horizon radius depends both on the parameters $r_0$ and $a_{01}$, this last condition also implicitly depends on both the $r_0$ and $a_{01}$ parameters.

Finally, in order to ensure that there exist no closed timelike curves outside the horizon requires that $g_{\phi\phi}>0$~\cite{Johannsen2013a}. This is always true for a spherically symmetric RZ spacetime but translates to the requirement $K^2>0$ in the presence of spin. Note that, although two of the deformation functions ($N^2$ and $K^2$) are written as perfect squares, their actual functional forms allow them to become negative, depending on the values and signs of the various deviation parameters.

Exploring the details of the event horizon for the general form of the RZ metric without spherical symmetry is beyond the scope of this paper. We note, however, that the shape of the horizon and its size will depend also on the spin $a$ as well as on the values of additional parameters such as $k_{21}$, $a_{20}$, and $a_{21}$. In fact, the shape of the horizon may be oblate or prolate in Boyer-Lindquist like coordinates, depending on the signs of these parameters. Furthermore, because the radius of the horizon  is not necessarily equal to $r_0$, the condition for a spherically symmetric horizon is not $k_{21}+a_{20}+a_{21}=0$ as suggested in Ref.~\cite{Konoplya2016}. Requiring that the horizon completely surrounds the central singularities and that the metric remains Lorentzian outside the horizon will introduce a number of additional bounds on the deviation parameters, which can be found following the procedures we outlined above. Such bounds on the deviation parameters have been explored in Ref.~\cite{Shashank2021}, albeit without using the correct location of the event horizon.

\section{Black-Hole Metrics From Modified Gravity Theories}

In this section, we evaluate the circularity condition for two metrics that are solutions to modified  field equations.

\subsection{Black-Hole Metrics in  Einstein-Maxwell-Dilaton-Axion Theories}

The first metric we investigate in this category is the Einstein-Maxwell-Dilaton-Axion metric. This describes a gravitional/electromagnetic system with the two additional fields, the scalar dilaton, which couples to the electromagnetic field via the $F_{\mu\nu} F^{\mu\nu}$, and the pseudoscalar axion, which couples to the electromagnetic field via the pseudoscalar $F_{\mu\nu} *F^{\mu\nu}$. 

We follow the notation of Ref.~\cite{Garcia1995}, but in order to be consistent with Ref.~\cite{Younsi2021}, we rename $\Sigma$ to $\hat{\Delta}$, $\Delta$ to $\hat{\Sigma}$, interchange $b$ and $\beta$ and finally replace $m$ by $m=M+b$. We then go to rational polynomial coordinates and set $G=c=M=1$ such that
\begin{equation}
ds^2 = -\frac{\hat{\Delta}-a^2 (1-z^2)}{\hat{\Sigma}} dt^2 - \frac{2 a (\hat{\delta}-\hat{\Delta} W) (1-z^2)}{\hat{\Sigma}} dt d\phi + \frac{\hat{\Sigma}}{\hat{\Delta}} dr^2 + \frac{\hat{\Sigma}}{(1-z^2)} dz^2 + \frac{\hat{A} (1-z^2)}{\hat{\Sigma}} d\phi^2
\end{equation}
with
\begin{eqnarray}
\hat{\Sigma} &=& r^2 + a^2 z^2 - (\beta^2 + 2 b r) + \ \beta_b \ (\beta_b -2 a z) \\
\hat{\Delta} &=& r^2 - 2 r + a^2 - (\beta^2 + 2 b r) - \ \beta_b^2 \ (1 + 2b)\\
W &=& 1 + \frac{\beta_{ab} (2 z - \beta_{ab}) + \beta_a^2}{1-z^2} \\
\hat{\delta} &=& r^2 - 2 b r + a^2 \\
\hat{A} &=& \hat{\delta}^2 -a^2 \hat{\Delta} W^2 (1-z^2)\;. 
\end{eqnarray}
Here, $b$ is the coupling parameter of the dilaton and $\beta$ the coupling parameter of the axion. The constants are defined as
\begin{eqnarray}
\beta_a = \frac{\beta}{a}\;, \ \ \ \ \ \ \beta_b = \frac{\beta}{b}\;, \ \ \ \ \ \ \beta_{ab} = \frac{\beta}{b a}\;.
\end{eqnarray}

The metric has a clear coordinate singularity at $\hat{\Delta}=0$, which has  been tentatively identified with the event horizon in earlier studies. We will show here that this condition indeed specifies the location  of the  event horizon, albeit for a different physical reason. In the following, we show explicitly the dependence of the various results on the mass $M$, for reasons that will become readily apparent.

The metric is clearly circular, as it is in the canonical form and the metric coefficients depend only on the coordinates $r$ and $z$. Morever, the determinant of the EMDA metric in rational polynomial coordinates is
\begin{equation}
\det g_{\mu\nu} = - (r^2 +a^2 z^2 -\beta^2- 2 b r - 2a z \beta_b + \beta_b^2 )^2 = -\hat{\Sigma}^2\;.   
\end{equation}
This shows that the EMDA metric is Lorentzian outside the singularity at $\hat{\Sigma} = 0$. Expanding $g_{tt}$ and $g_{t\phi}$ in Boyer-Lindquist coordinates shows that the metric in the dilaton-only case, with the axion coupling constant $\beta$ set to zero, is asymptotically flat. The metric with $\beta \ne 0$ is not asymptotically flat, as the series expansion of $g_{t\phi}$ in the radial coordinate contains a constant term proportional to $\beta$, which is not admissible for asymptotically flat spacetime \cite{Heusler1996}. This is not surprising, as the axion parameter $\beta$ is related to the NUT parameter of the asymptotically non-flat Taub-NUT spacetime~\cite{Garcia1995}.

This allows us to apply Carter's theorem (albeit with constraints on the various coupling parameters to ensure that $\hat{\Sigma}\ne 0$) only in the case with $b \ne 0$ and $\beta = 0$. Nonetheless, we perform the following calculations for the general case $b, \beta \ne 0$. We write the determinant of the $t$, $\phi$ part of the  metric as
\begin{equation}
{\cal N} =   - (a^2 -2 r +r^2 +\beta^2 -2br + 2 b \beta_b^2 - \beta_b^2) (1-z^2) = - \hat{\Delta}
\end{equation}
and conclude (at least in the case of $\beta = 0$) that, indeed, the event horizon in the EMDA metric is located at $\hat{\Delta} = 0$, i.e., at
\begin{eqnarray}
r_{\rm h} = (1+b) + \sqrt{(1+b)^2 (1+\beta_b^2)-a^2} \;. 
\end{eqnarray}
It is through the application of the weak rigidity and rotosurface theorems that we can identify this location with the event horizon of the spacetime, even though it coincides with the location of the coordinate singularity. It is also interesting that both the dilaton and the axion modify the mass-dependent term of the horizon (when compared to the Kerr value) but not the spin-dependent term. 

As implied by the rigidity theorem, the angular velocity is constant on the outer event horizon, i.e.,
\begin{equation}
\Omega_H = \frac{a}{\hat{\delta}} = \frac{a}{r_+^2 +a^2 -2b r_+} = \frac{a}{2(1+b)r_+}\;.
\end{equation}
The surface gravity on the horizon with no axions ($\beta=0$) is
\begin{equation}
\kappa = \frac{r_{\rm +} -(1+b)}{2 r_{\rm +}},
\end{equation}
whereas with no dilaton ($b=0$) it is
\begin{equation}
\kappa = \frac{r_{\rm +} -1}{2 r_{\rm +} + \beta^2}\;.
\end{equation}
Note that in this metric, the sum $\Omega_H^2+\kappa^2$ depends not only on $r_+$ but also on the spin $a$ and the dilaton parameter $b$.

\subsection{Black Hole Metrics in Degenerate Higher Order Scalar Tensor (DHOST) Theories}

As a final example, we contrast the results presented earlier with those of a class of black-hole metrics in DHOST theories~\cite{Anson2021,Achour2020}. These are stationary axisymmetric metrics that are not Ricci flat. However, because the underlying scalar field that sources them is time evolving, the resulting metric contains off-diagonal terms and cannot be put in the Lewis-Papapetrou form.

An example DHOST black-hole metric is
\begin{eqnarray}
ds^2 &=& -\left(1-\frac{2\tilde{M}r}{\Sigma}\right) dt^2 - \sqrt{1+D} \frac{4 \tilde{M} a r \sin^2 \theta}{\Sigma} dt d\phi+ \frac{A \sin^2 \theta}{\Sigma} d\phi^2 + \nonumber \\ & & \left[\frac{\Sigma}{\Delta}-D(1+D)\frac{2\tilde{M}r(r^2+a^2)}{\Delta^2}\right] dr^2-  2D\frac{\sqrt{2\tilde{M}r(r^2+a^2)}}{\Delta}dr dt + \Sigma d\theta^2
\end{eqnarray}
with $\Delta(r) = r^2 - 2 M r +a^2 $,  $\Sigma(r,\theta) = r^2+a^2 \cos^2 \theta $ and $ A(r,\theta) = ( r^2+a^2)^2 - a^2 \Delta \sin^2 \theta$. The disformal parameter $D$, which is associated with the linear rate of change of the scalar field, sets the mass and spin of the black hole, as measured by an observer at infinity, to $\tilde{M} = M/(1+D)$ and $\tilde{a}=a\sqrt{1+D}$, respectively. The limit $D=0$ reproduces the standard Kerr solution. 

The DHOST metric is asymptotically flat and admits an asymptotically timelike Killing vector k (=$\partial_t$) and a second Killing vector m ($=\partial_{\phi}$) for axisymmetry. However, as discussed extensively in Ref.~\cite{Anson2021}, the presence of these two symmetries does not guarantee that the circularity condition is satisfied. Indeed, the metric contains a $g_{tr}$-term, which arises from the time-evolution of the scalar field. As a result, the 1-form associated to the time-like Killing vector is given by $k=g_{tt}dt + g_{t\phi} d\phi + g_{tr} dr$, whereas the 1-form associated to the spatial Killing vector remains $m = g_{\phi t} dt + g_{\phi\phi} d\phi$. This leads to the following equalities:
\begin{eqnarray}
m \wedge k \wedge dk &=& g_{tr} [-(\partial_{\theta} g_{tt}) g_{\phi\phi} + (\partial_{\theta} g_{t\phi}) g_{t\phi}] \ dt \wedge \ dr \wedge \ d\theta \wedge d\phi \\
k \wedge m \wedge dm &=& g_{tr} [-(\partial_{\theta} g_{\phi\phi}) g_{t\phi} + (\partial_{\theta} g_{t\phi}) g_{\phi\phi}] \ dt \wedge \ dr \wedge \ d\theta \wedge d\phi \;.
\end{eqnarray}
Because $g_{tr}$ is non-zero and the $tt-$, $\phi\phi$-, and $t\phi$-components of them metric are $\theta$-dependent, the above two expressions are non-zero and, therefore, the metric is not circular. The only situation in which the metric is circular is when $\partial_\theta g_{tt}=\partial_\theta g_{\phi\phi}=\partial_\theta g_{t\phi}=0$. However, this is equivalent, with an appropriate redefinition of parameters, to the Schwarzschild metric and, hence,  does not represent a new black-hole solution.

Ref.~\cite{Anson2021} explores in detail the properties of this metric and offers a possible calculation of its event horizon. For the purposes of our paper, we present it only as a counterexample of a black-hole spacetime that is stationary and axisymmetric but, because it is not circular, does not allow us to use the weak rigidity theorem and Carter's rotosurface theorem to identify its event horizon.

\section{Conclusions}

We have shown that the application of Carter's rotosurface theorem for circular metrics allows us to analytically calculate the event horizon for several of the most commonly used non-GR metrics, as they tend to be circular. We pointed out instances where the application of the theorem is not possible, since some of its assumptions, like asymptotic flatness, are not satisfied. We also discussed non-circular non-GR metrics, for which other methods to determine the event horizon have to be devised.

Our approach allowed us to revisit the investigation of the event horizon location in the commonly used RZ metric and also to find a discontinuous jump in the dependence of this location on particular deviation parameters. Identifying the correct location of the event horizon is important in setting bounds on the deviation parameters such that the parametric metric is free of pathologies.

The application of the weak rigidity and rotosurface theorems allowed us also to calculate the surface gravity of the horizon and its angular velocity. Perhaps not surprisingly, these two properties appear to be directly related to the predicted shape and size of the predicted black-hole shadows for these metrics. For the JP metric, we found that the surface gravity of the horizon depends on the same deviation parameters that affect the size of the black-hole shadow (e.g., $\alpha_{13}$), whereas the angular velocity of the horizon depends on the same deviation parameters that affects the shadows shape and, in particular, any deviation from circular symmetry (e.g., $\alpha_{22}$)~\cite{Johannsen2013c,Medeiros2020,Psaltis2020}. This is easy to understand since the effective gravity near the horizon affects the radius of the photon orbit and, hence, the size of the black-hole shadow. On the other hand, the angular momentum of the horizon is a measure of frame dragging, which is responsible for any deviations of the shadow shape from circular. In a future publication, we will explore this potential connection to a greater extend. 

\bigskip

{\em Note added.---\/} During the final stages of our work, we became aware of Ref.~\cite{Delaporte2022}, which proposes a new class of parametric metrics that deviate from Kerr that allow for violation of the circularity condition. 

\begin{acknowledgments}
We are grateful to F.\ \"Ozel and P.\ Christian for many discussions and for comments on the manuscript. This work was supported in part by NSF PIRE award OISE-1743747, NSF award AST-1715061, and NASA ATP award 80NSSC20K0521.
\end{acknowledgments}

\bibliography{circular.bib}

\end{document}